
\magnification = \magstep 1
\openup 2\jot

\centerline {LINEAL GRAVITY FROM PLANAR GRAVITY}

\vskip 1 truecm
\centerline { Ana Ach\'ucarro}
\vskip 1 truecm

\centerline {Department of Mathematics}
\centerline  {Tufts University, Medford, MA 02174}

\centerline {and}

\centerline {Department of Theoretical Physics}
\centerline {University of the Basque
Country (Spain)} \vskip 5 truecm
\centerline {Abstract}

We show how to obtain the two dimensional black hole action by a dimensional
reduction of the three dimensional Einstein action with a non-zero cosmological
constant. Starting from the Chern-Simons formulation of 2+1 gravity, we obtain
the 1+1-dimensional gauge formulation given by Verlinde. Remarkably, the
proposed reduction shares the relevant features of the formulation  of Cangemi
and Jackiw, without the need for a central charge in the
algebra. We show how the Lagrange multipliers in these
formulations appear naturally as the remnants of the three-dimensional
connection associated to symmetries that have been  lost in the dimensional
reduction. The proposed dimensional reduction involves a shift in the three
dimensional connection whose effect is to make the length of the extra
dimension
infinite.

\vfill \eject

The past few months have seen a revival  in the study of the quantum
properties of black holes due to the discovery that relatively simple actions
in
two dimensions admit  black hole solutions [1,2]. These are being
intensively studied as toy models in which to investigate back-reaction
in Hawking radiation and related issues in quantum gravity,  in particular the
formation of black holes and the final stages of black hole evaporation [2,3].
The hope is to find a consistent quantization scheme for 2d  gravity coupled
to  matter that will answer some of the  questions that are intractable in four
dimensions.

 We  expect the underlying group
structure to be crucial in the quantization process.
 Exact quantization of
three-dimensional gravity was achieved by Witten using the fact that it can be
rewritten as a Chern-Simons action for the tangent space group [4,5].
The two-dimensional ``Einstein'' action is a topological invariant
and therefore has trivial dynamics. Alternatives include the action proposed
by Teitelboim [6] and Jackiw [7] and the above mentioned
 `string-inspired'
 action [2,8]. The gauge-theoretical formulation of the first action [6,7],
based on the group $SO(2,2)$, was
obtained by Chamseddine and Wyler [9] and by Isler and Trugenberger [10]. The
group formulation of the
`string-inspired' action  has recently been found in two very interesting
papers, which offer different  answers.  The formulation proposed by Verlinde
[8] is loosely based on the group $ISO(1,1)$. That of
Cangemi and Jackiw [11] is, in some sense, more ``natural'' - but at
the expense of introducing a central charge in the algebra.

A general feature of
all these two-dimensional actions is that their gauge theoretical  formulations
 include extra fields  (Lagrange multipliers) which do not come from
the metric, and whose geometric interpretation is unclear. In this paper we
find a geometric interpretation of these Lagrange multipliers by a process of
dimensional reduction  from pure gravity in three  dimensions. They are
remnants of the three-dimensional theory associated with the generators that
disappear in the reduction. In the process, we recover Verlinde's formulation
of the `black hole' action, but in a way that shares  the good features of the
formulation of Cangemi and Jackiw.
Our starting point is pure 2+1 dimensional gravity (with a cosmological
constant) in its Chern-Simons formulation.  For completeness, and because
it illustrates  the main features of the reduction very clearly, we first
obtain the action of [6,7] by a perfectly standard
dimensional reduction. We then propose a non-standard reduction scheme
involving a shift in the three dimensional connection, which yields the
`black hole'   action of [2,8].

We begin by considering  the Einstein action
for 2+1 dimensional gravity with a cosmological
constant
$$
S = \int d^3x \sqrt{g} (R_g - 2\Lambda)
\eqno (1)
$$
where $g$ is the determinant of the metric $g_{IJ}$ ($I,J = 0,1,2$), and $R_g$
is the Ricci scalar. $(- 4\Lambda)$ is the cosmological constant; in what
follows we shall take it to be negative, i.e. $\Lambda >0$, but the same
analysis goes through for $\Lambda <0$.
Our  conventions are as follows:
we will use capitals for three-dimensional indices and lowercase for
two-dimensional ones. $I, J,  i,j...$ are spacetime indices, while $A,B,a,b...$
are tangent space indices.
The former are raised and lowered with the three dimensional metric $g_{IJ}$
or the two-dimensional metric $\gamma_{ij}$, the latter with the Minkowski
metrics  $\eta_{AB} = (+1, -1, -1)$ or $\eta_{ab} =
(+1, -1)$. $\epsilon_{ABC}$ is the totally antisymmetric
symbol, with $\epsilon^{012} = 1$. In two dimensions, $\epsilon^{ab} =
\epsilon^{ab2}$, so $\epsilon^{01} = 1$ also.

One can also formulate the action (1) in what is called a first order
formalism. Given coordinates $(x^I)$,
the idea is to introduce  an orthonormal triad $\{ \vec{e_A} \}$,
with components ${\vec e_A} = e_A^{\ \ I} \partial / \partial x^I$ in the
coordinate basis,  and a spin connection $\{\omega_I^{\ \ A} \}$ such that the
torsion constraints
$$\epsilon^{IJK} (\partial_J e_K^{\ \ A} - \epsilon^A_{\ \ BC}
\omega_J^{\ \  B} e_K^{\ \  C} ) = 0
\eqno (2)$$ are satisfied.
The $e_I^{\ \ A}$ are defined as the inverses of the $e_A^{\ \ I}$, thus
$$e_I^{\ \ A} e_A^{\ \ J} = \delta _I^{\ \ J}\ \ \ \ \ \ \ \ \ \ \ \ \ \
e_A^{\ \ I} e_I^{\ \ B} = \delta _A^{\ \ B}\eqno(3)$$
and the metric is obtained from
 $$g_{ij} = e_I^{\ \ A} e_J^{\ \ B}
\eta_{AB}\eqno(4)$$
In particular, the three-dimensional Riemann tensor can be written
$$
R_{IJKL} = \epsilon_{ABC} e_K^{\ \ A} e_L^{\ \ B}
[ \partial_I \omega_J^{\ \ C} - {1 \over 2} \epsilon^C_{\ \ DF} \omega_I^{\ \
D}
\omega_J^{\ \ F} ] \ - \ (I \rightarrow J)
\eqno (5)$$
and the equations of motion of (1) reduce to
$$\partial_I \omega_J^{\ \ A} -{1 \over 2} \epsilon^A_{\ \ BC} \omega_I^{\ \ B}
\omega_J^{\ \ C} - {\Lambda \over 2} \epsilon^A_{\ \ BC} e_I^{\ \ B} e_J^{\ \
C}
 \ - \ (I \rightarrow J) = 0
\eqno (6)
$$

The second order formulation is recovered by
solving the torsion constraints (2) for the $\omega$'s as a function of
the $e$'s and substituting $\omega(e)$ in (6).

We can rewrite the equations in a coordinate free way by introducing the
one-forms $e^A$ and $\omega^A$ (here, $e^A = dx^I e_I^{\ \
A}$ $\omega^A = dx^I \omega_I^{\ \ A}$). We then have
$$
\eqalign{
T^A &= De^A = de^A - \epsilon^A_{\ \ BC} \omega^B e^C = 0  \ \cr
R^A &= d\omega^A - {1 \over 2} \epsilon^A_{\ \ BC} \omega^B \omega^C -
{\Lambda \over 2} \epsilon^A_{\ \ BC} e^B e^C = 0 \cr}
\eqno (7)
$$
If we introduce a  $SO(2,2)$ connection
$$
{\rm A} = e^A P_A + \omega^A J_A \ \ ,
\eqno (8)$$
where $P_A$ and $J_A$ are translations and Lorentz rotations,
satisfying
 $$
[P_A, P_B] = -\Lambda  \epsilon_{ABC}J^C \ \ \ \ \ \ \ \ \
[J_A, P_B] = -\epsilon_{ABC} P^C \ \ \ \ \ \ \ \ \
[J_A, J_B] = -\epsilon_{ABC} J^C \ \ ,
\eqno (9)$$
the equations (7) are immediately recognized as the components of the
curvature two-form of $SO(2,2)$
$$
{\rm F} = d {\rm A} + {1 \over 2}[{\rm A}, {\rm A}] = T^A P_A + R^A J_A
\eqno (10)$$
and the first order action corresponding to (1) turns out to be the
Chern-Simons action for this connection [4,5].
 $$
S = \int {\rm tr}( {\rm A} d{\rm A} + {2 \over 3} {\rm A}^3 )
\eqno (11)
$$
where the exterior product of differential forms is understood. (The same is
true for
$SO(3,1)$ if the cosmological constant is positive,  or for $ISO(2,1)$ if
it is zero).

 The ``trace"  in the action refers to any invariant bilinear form in
the  algebra, and in this case it is given by the Casimir
$$
 C =  P_A J_B ~\eta^{AB}
\eqno (12)$$
Note that the  algebra is semisimple, $so(2,2) = so(2,1) + so(2,1)$, with the
two factors  generated by $M^\pm_A = {1 \over 2} (J_A \pm P_A/\sqrt{\Lambda})
$.
This means that there is a one-parameter family of Casimirs (up to overall
normalization) $$
C(\sigma) = M^+_A M^{+A} \ + \ \sigma M^-_AM^{-A}
\eqno (13)$$
All the actions obtained for different values of $\sigma$ are classically
equivalent in the sense that their equations of motion are the same,
$\ {\rm F}^+ = {\rm F}^- = 0~$. The Einstein action is obtained for $\sigma =
-1$

{}From now on, the indices a,b will take the values 0,1 only, and we will
indicate the index 2 explicitly.

The first action that we want to investigate was proposed by Teitelboim
[6] and Jackiw [7]
 $$
S = \int d^2x \sqrt{-  \gamma} \Phi (R_{\gamma} - 2\Lambda)
\eqno (14)
$$
($\gamma = {\rm det} \gamma_{ij}$). $\Phi$ is a Lagrange multiplier
enforcing the constraint
$$
R_{\gamma} - 2\Lambda = 0
\eqno (15)$$

In order to obtain this action in the second order formulation by
dimensional reduction of the three dimensional Einstein action, we consider
coordinates $(x^I) = (x^i, y)$, we compactify the $y$ coordinate by
identifying the points whose $y$ coordinate differs by L and we impose that
all derivatives of the metric in the y direction be zero. The condition
$\partial_y g_{IJ} = 0 $ is invariant under two-dimensional diffeomorphisms $
x^i \rightarrow x'^i(x^j) $ and also under the residual transformation $y
\rightarrow y + f(x^i) $, where $f$ is an arbitrary function. If we parametrise
the three-dimensional metric as
 $$
g_{IJ} = \pmatrix {\gamma_{ij} - A_iA_j  \ \ \  \Phi A_i \cr \  \Phi A_j \ \
 \ \ \ \ \ \ \ \  -\Phi^2 \cr } \ \ ,
\eqno (16)$$
the effect of this last transformation on the metric is,
infinitesimally,
 $$
\delta \Phi = 0 \ \ \ \ \ \ \ \ \ \
\delta A_i = \partial_i f\ \ \ \ \ \ \ \ \ \
\delta \gamma_{ij} = 0 \ \ \ \ \ \ \ \ \ \
\eqno (17)$$
and therefore we can use it to set $A_i = 0$. In that case, $\sqrt {g} =
\Phi \sqrt {- \gamma}$, and (1) yields (14) plus a total derivative. This is
the
dimensional reduction that was proposed in [7].

Dimensional reduction is also straightforward in the first
order formulation.
Starting with (11) one obtains the gauge-theoretical formulation of
(14)   given by Isler and Trugenberger [10], and by Chamseddine and
Wyler [9]. They introduce a gauge connection for the anti-de Sitter group
$SO(1,2)$  $${\rm A} = e^a P_a + \omega J \eqno (18)$$ and a triplet of
Lagrange
multipliers $\chi_A$ transforming under the coadjoint representation and write
the action as  $$
\eqalign{
S &= \int \chi_A F^A   \cr
&= \int \bigl[ \chi_a (de^a + \epsilon^{ab} \omega e_b) + \chi_2 (d \omega +
{\Lambda \over 2} \epsilon_{ab} e^a e^b) \bigr] \cr}\eqno (19)$$
where $F^A$ are the components of the curvature two-form,
$$F = dA + {1 \over 2}[A,A] = (de^a + \epsilon^{ab}\omega e_b ) P_a + (d\omega
+
{\Lambda \over 2}\epsilon_{ab} e^a e^b) J \eqno (20)$$
with
$$[P_a, P_b] = \Lambda \epsilon_{ab} J \ \ \ \  \ \ \ \ \ \
[P_a, J] = \epsilon_{ab} P^b
\eqno (21)
$$

In order to obtain this formulation by dimensional reduction, we need an
ansatz for the $e^A, \omega^A$ equivalent to (16). Since
$$
g_{IJ}= e_I^{\ \ a} e_J^{\ \ b} \eta_{ab} - e_I^{\ \  2} e_J^{\ \ 2} \ \ ,
\eqno (22)$$
the dreibein components are independent of $y$ :
$$
e_y^{\ 2} = \Phi (x)\ \ \  \ \  \ \ \ \ \ \  \
e_i^{\ 2} = A_i (x)\ \ \ \ \ \ \ \ \ \ \ \ \ \
e_i^{\ a} e_j^{\ b} \eta_{ab} = \gamma_{ij} (x)
\eqno (23)$$
Setting $A_i = 0$, the connection one-forms become
$$
e^a = dx^i e_i^{\ a}  \ \ \ \ \ \ \ \ \
e^2 = dy ~\Phi
\eqno (24)$$

We now look at the ansatz for the spin connection.
Any ansatz used in the dimensional reduction must be consistent with the
equations of motion of the three-dimensional theory. Introducing $\omega^A =
dx^i \omega_i^{\ A} + dy ~ \omega_y^{ \ A}$ in the equations (7), we find
that consistency requires $
\omega_i^{\ a} =  \omega_y^{\ 2} = 0 $, so the correct ansatz is
$$
\omega^a = dy~\omega_y^{\ a} (x) \ \ \ \ \ \ \ \ \ \ \ \ \ \ \ \ \ \
\omega^2 = dx^i \omega_i^{\ 2}
\eqno (25)$$

 We now perform a rescaling of the algebra
generators $ J_a \rightarrow J_a / \mu , \
P_2 \rightarrow P_2 / \mu$.
The  commutation relations (9) become
$$
[P_a, P_2] =  \Lambda \epsilon_{ab} J^b \ \ \ \ \ \ \ \ \ \  \ \ \ \ \ \
[J_a, P_2] = \mu^2\epsilon_{ab} P^b \ \ \ \ \ \ \ \ \ \ \ \ \ \ \ \
[J_a, J_b] = \mu^2\epsilon_{ab} J_2
$$
$$
[J_a, P_b] = \epsilon_{ab} P_2 \ \ \ \ \ \ \ \ \ \ \
[J_a, J_2] = \epsilon_{ab} J^b
\eqno (26)$$
together with those of the two-dimensional anti-de Sitter group $SO(2,1)$
$$
[P_a, P_b] = \Lambda \epsilon_{ab} J_2 \ \ \ \ \ \ \ \ \ \
[J_2, P_a] = -\epsilon_{ab} P^b \ \ \ \ \
\eqno (27)$$
{}From now on, $J_2 = J$, the two dimensional Lorentz generator, and $\omega^2
=
\omega$. The rescaled Casimir  $$ \mu C =  P_a J^a - P_2 J \eqno (28)$$
survives in the limit $\mu \rightarrow 0$, and the action (11) becomes
$$
S =  \int dy \int d^2x [\omega_{ya} (de^a + \epsilon^{ab} \omega e_b)
- \Phi (d\omega + {\Lambda \over 2} \epsilon_{ab} e^a e^b )`] \ \ .
\eqno (29)$$
Integration over $y$ yields a factor of L at the front, and we obtain the
action (19). Notice that  the Lagrange multipliers $\chi^a = \omega_y^{\ a},
\ \chi^2 = e_y^{\ 2} = \Phi$ appear naturally; they are  simply the components
of the three-dimensional connection corresponding to the generators $ J_0,\
J_1$  and $P_2$ that have been lost in the two dimensional theory.

We now turn to the `string-inspired' model of [2], or rather to that
of [8] ,
 $$
S_2 = \int d^2x \sqrt{-\gamma}(\Phi R_{\gamma} - \lambda) \ \ ,
\eqno (30)
$$
which is
obtained from the  action of [2] by a conformal transformation.

There are two ways of writing this action in a gauge-theoretical framework.
The first one, proposed by Verlinde [8] uses the $ISO(1,1)$ group
$$
[P_a, P_b] = 0 \ \ \ \ \ \ \ \ \ \ \ \ \ \ \ \ \ [P_a, J] = \epsilon_{ab} P^b
\eqno (31)$$
The action is written as
$$
\eqalign{
S_2 &= \int  (\chi_A {\rm F}^A - \lambda\epsilon_{ab} e^a e^b )\cr
    &= \int \bigl[ \chi_a (de^a + \epsilon^{ab} \omega e_b) + \chi_2 d \omega
           - \lambda\epsilon_{ab} e^a e^b \bigr] \cr}
\eqno (32)
$$
and it is invariant under the  transformations
$$
\eqalign{
\delta e^a &= d \theta^a +  \epsilon^{ab}\omega \theta_b + \epsilon^{ab}
\alpha  e_b \cr
\delta \omega \ &= d \alpha                            \cr
\delta \chi^a &= \epsilon^{ab} \alpha \chi_b + \lambda \epsilon^{ab} \theta_b
  \cr
\delta \chi^2 &=  \epsilon_{ab} \chi^a \theta^b
\cr}  \eqno (33)$$

The second one, proposed by Cangemi and Jackiw, uses a central extension of
the $ISO(1,1)$ algebra
$$
[P_a, P_b] = {\lambda }\epsilon_{ab}  I\ \ \ \ \ \ \ \ \ \ \ \ \ \
\ [P_a, J] = \epsilon_{ab} P^b \ \ \ \ \ \ \ \ \ \ \  \ \ \ \ \ [I,
P_a] = [I,J] = 0
\eqno (34)$$ which has a non-degenerate Casimir
$$
C = P_aP^a - {\lambda } I J \ \ \ .
\eqno (35)$$
Introducing ${\rm A} = e^aP_a + \omega J + {\lambda } a I$, the action
becomes
$$
\eqalign{
S_2 &= \int \chi_A {\rm F}^A \cr
    &= \int \chi_a (de^a + \epsilon^{ab} \omega e_b) + \chi_2 d \omega
           + \chi_3 (da + {1 \over 2} \epsilon_{ab} e^a e^b)\cr}
\eqno (36)
$$
(where now the index $A$ runs from 0 to 3) and is invariant under the natural
gauge transformations $$\delta {\rm A} = d {\rm A} + [{\rm A} , {\rm v}]
\eqno (37)$$
for a gauge parameter ${\rm v} = v^a P_a + v J +  \lambda u I$.
In this case, $${\rm F} = (de^a + \epsilon^{ab} \omega e_b) P_a + d\omega J +
{\lambda } (da + {1 \over 2} \epsilon_{ab} e^a e^b ) I
\eqno (38)$$
Note that there are {\it four} Lagrange multipliers. Verlinde's formulation is
recovered after elimination of $a$ and $\chi_3$ by their equations of motion,
notably $d\chi_3 = 0$, which allows us to set $\chi_3 = -2\lambda$.

To obtain a 2d formulation starting wiht the 3d Chern-Simons action, we
introduce a different rescaling of the $SO(2,2)$ generators, which now includes
a rescaling of the $P_a$. This is just the usual rescaling  in Wigner-In\"onu
contraction that obtains the underlying Poincar\'e group  $ISO(1,1)$ in the
limit where $\Lambda \rightarrow 0$.
(Note that $J_2$ cannot be rescaled if it is to generate
2-dimensional Lorentz rotations, since a rescaling would alter its commutation
relation with  translations).
The rescalings $$ P_a \rightarrow {P_a \over
\sqrt{\Lambda}} \ \ \ \ \ \ \ \ \ \ \ \ \  P_2 \rightarrow {P_2 \over \mu
\sqrt{\Lambda}} \ \ \ \ \ \ \ \ \ \ \ \ \ J_a \rightarrow {J_a \over \mu} \ \ \
\ \ \ \
\eqno (39)$$
together with the constant shift
$$
e^a \rightarrow e^a + {\lambda \over \Lambda} dy
\eqno (40)$$
give the action (30) in the limit where $\mu, \Lambda \rightarrow 0$
provided $\mu^2 / \Lambda \rightarrow 0$. The rescaling of $P_2$ is determined
by the condition that the Casimir be non-degenerate after the contraction.

The commutation relations (9) now become
$$
[P_a, P_2] = \Lambda \epsilon_{ab} J^b \ \ \ \ \ \ \ \ \
[J_a, P_2] = \mu^2 \epsilon_{ab} P^b \ \ \ \ \ \ \ \ \
[J_a, J_b] = \mu^2 \epsilon_{ab} J \ \ \ \ \ \ \ \ \
$$
$$[J, J_a] = -\epsilon_{ab} J^b \ \ \ \ \ \ \ \ \
[J_a, P_b] =  \epsilon_{ab} P_2
\eqno (41)$$
and those of the two-dimensional theory,
$$
[P_a, P_b] = \Lambda \epsilon_{ab} J \ \ \ \ \ \ \ \ \
[J, P_a] = -\epsilon_{ab} P^b
\eqno (42)$$
Note that in the limit $\mu = \Lambda = 0$, the algebra is invariant under the
interchange of $J_a$ and $P_a$. As a consequence,
the rescaled Casimir
$$
 (\mu \sqrt{\Lambda} )C =  P_a J^a - P_2 J
\eqno (43)$$
is essentially equivalent to that appearing in
the $ISO(1,1)$ algebra with central charge, proposed  by Cangemi and Jackiw
(even though $P_2$ is {\it not} a central charge). The first term behaves like
$P_aP^a$ and the second term plays the role of the $JI$ term in [11],
making the Casimir non-degenerate. Note that there is no rescaling that will
give the central charge in the algebra while keeping J as the generator
of Lorentz rotations.

With this ansatz, the
Chern-Simons  action becomes $$
S =  \int dy \int \biggl[
\omega_{ya} (de^a + \epsilon^{ab} \omega e_b) - \Phi  d\omega  - \lambda
\epsilon_{ab} e^a e^b \biggr] \ + \ ...
\eqno (44)$$
where the ... indicates terms that vanish in the limit $\mu, \Lambda
\rightarrow 0$. This is precisely (32) and again we recognise the Lagrange
multipliers as coming from $\omega^a$ and $e^2$.

The gauge transformations $\delta A = dv + [A,v]$ with
$$\eqalign{
{\rm A} &= e^a P_a + \omega J + \omega^a J_a + e^2 P_2 \cr
v &= \theta^a P_a +   \alpha J + \beta^a J_a + \rho P_2 \cr}
\eqno (45)$$
become
$$
\eqalign{
\delta e^a &= d \theta^a +  \epsilon^{ab}\omega^2 \theta_b + \epsilon^{ab}
\alpha e_b + \mu^2 \epsilon^{ab} (e^2 + {\lambda \over \Lambda} dy ) \beta_b
+ \mu^2 \epsilon^{ab} \rho \omega_b  \cr
\delta \omega^2 &= d \alpha + \mu^2 \epsilon_{ab}\omega^a \beta^b   +
\Lambda \epsilon_{ab} e^a \theta^b                            \cr
\delta \omega^a &= d\beta^a + \epsilon^{ab}\alpha \omega_b +
\epsilon^{ab} \omega^2 \beta_b + \Lambda \epsilon^{ab} \rho e_b
+ \Lambda (e^2 + {\lambda \over \Lambda}dy ) \epsilon^{ab}
\theta_b      \cr
\delta e^2 &= d\rho +  \epsilon_{ab} \omega^a \theta^b +
\epsilon_{ab} e^a \beta^b
\cr} \eqno (46) $$
In the limit when $\mu$ and $\Lambda$ tend to zero,  we
recover the transformations proposed by Verlinde by setting $\beta^a = \rho =
0$. The remaining  symmetries of the action change it
by  a total derivative, and only affect the Lagrange multipliers. They are
given by $$
\delta_{\beta} \omega^a = d \beta^a +\epsilon^{ab}\omega\beta_b \ \ \ \ \ \
\ \  \delta_{\beta} e^2 =  - \epsilon_{ab} e^a \beta^b
\eqno(47)$$
and
$$
\delta_{\rho} \omega^a = 0 \ \ \ \ \ \ \ \ \ \ \ \ \ \ \ \ \ \ \
\delta_{\rho} e^2 = d \rho
\eqno (48)$$

Note that the shift $e^2 \rightarrow e^2 + (\lambda / \Lambda ) ~dy$
corresponds
to a shift
$$\Phi^2 \rightarrow \bigl(\Phi + {\lambda \over \Lambda}\bigr)^2\eqno (49)$$
and it has the effect of putting all points in the y direction
an infinite distance away. This shift in $\Phi$ can be made directly in the
action (1), or in its dimensional reduction (14)
$$
S_2 = \lim_{\Lambda \rightarrow 0} \int d^2x \sqrt {-\gamma}(\Phi + {\lambda
\over \Lambda}) (R_{\gamma} - 2\Lambda)
\eqno (50)$$
(Incidentally, this clearly illustrates that in $S_1$ the two-dimensional
cosmological constant is the same as that of the three-dimensional theory,
whereas this is not so in the string-inspired action $S_2$).
Note that this procedure  would give rise to
divergent terms in the limit where $\Lambda$ goes to zero {\it in any dimension
other than two}. The reason why it is acceptable in two dimensions is that the
$1 / \Lambda$ term  in the action happens to be multiplying the Euler
characteristic, which is a topological invariant, and does not affect the
classical equations of motion of the two dimensional theory.  On the other
hand, the presence of such a term in the path integral merits investigation,
as it appears to  favour spheres, while supressing
higher genus topologies.

Of course the most interesting  question at the moment is the quantization of
2-dimensional gravity with matter. (Both
theories considered here have been  quantized succesfully in the absence of
matter [2,6-10], but the quantization   with matter seems a much harder
problem). There are a number of ways of  coupling matter to gravity in 2+1
dimensions using the underlying group structure (see, for example, [4,12])
and one might gain insight into this problem by  studying  their dimensional
reduction to two dimensions. Another interesting problem is to find  the three
dimensional solutions  which correspond to the 2d black holes. Finally, one
could investigate if there are other rescalings of the three dimensional
generators which yield interesting two dimensional actions.

\vskip 3 truecm
\noindent
NOTE: After completion of this work, we discovered a paper by Cangemi (we thank
Herman Verlinde for pointing it out) in which he proposes a different
dimensional reduction, starting from a 2+1 dimensional model based on an
abelian extension of the  Poincar\'e algebra with three extra generators [13].

\vskip 3 truecm
\noindent ACKNOWLEDGEMENTS

I would like to thank   Peter
Bowcock, Ruth Gregory, Mark Hindmarsh, Miguel Ortiz, John Stewart and Gerard
Watts for interesting conversations. I am also grateful to DAMTP for their
hospitality, and to the NSF for financial support.

\vskip 3 truecm
\noindent
REFERENCES

\noindent
[1] E. Witten, Phys. Rev. {\bf D44} (91) 314.

\noindent
[2] C. Callan, S. Giddins, J. Harvey and A. Strominger, Phys. Rev.{\bf D45}
(92) 1005

\noindent
[3] See, for instance, S.W. Hawking ``Evaporation of two-dimensional black
holes'', Caltech/DAMTP preprint CALT-68-1774; J.G.Russo , L. Susskind and L.
Thorlacius, ``The endpoint of Hawking  radiation'', Stanford Univ. preprint
SU-ITP-92-17; T. Banks and M. O'Loughlin, ``Classical and quantum producton of
cornucopions at energies below $10^{-18}$ GeV'', Rutgers preprint RU-92-14.
(And references therein).

\noindent
[4] A.Ach\'ucarro and P.K. Townsend, Phys. Lett. {\bf 180B} (86) 89; Phys.
Lett. {\bf 229B} (89) 383.

\noindent
[5] E. Witten, Nucl. Phys. {\bf B311} (88) 46; Nucl. Phys. {\bf B323}
(89) 113.

\noindent
[6] C. Teitelboim, Phys. Lett {\bf 126B} (83) 41

\noindent
[7] R. Jackiw, Nucl. Phys. {\bf B252} (85) 343

\noindent
[8] H. Verlinde, Proceedings of the Sixth Marcel Grossmann Meeting, to
be published.

\noindent
[9] A. Chamseddine and D. Wyler, Phys. Lett. {\bf 228B} (89) 75

\noindent
[10] K. Isler and C. Trugenberger, Phys. Rev. Lett {\bf 63} (89) 834

\noindent
[11] D. Cangemi and R. Jackiw, ``Gauge invariant formulation of lineal
gravity'', MIT preprint CTP 2085

\noindent
[12] M.E. Ortiz, Nucl. Phys. {\bf B363} (91) 185.

\noindent
[13] D. Cangemi, ``One formulation for both lineal gravities through a
dimensional reduction'',  MIT preprint CTP 2124

 \bye